\documentclass[pre,twocolumn]{revtex4}
\usepackage[english]{babel}
\usepackage{amsmath}
\usepackage{amssymb}
\usepackage{epsfig}

\begin{document}

\title{Discrete Vortices in Systems of Coupled Nonlinear Oscillators:\\
Numerical Results for an Electric Model}
\author{Victor P. Ruban}
\email{ruban@itp.ac.ru}
\affiliation {Landau Institute for Theoretical Physics, Russian Academy of Sciences,
Chernogolovka, Moscow region, 142432 Russia}

\date{\today}

\begin{abstract}
Vortex coherent structures on arrays of nonlinear 
oscillators joined by weak links into topologically nontrivial
two-dimensional discrete manifolds have been theoretically 
studied. A circuit of nonlinear electric oscillators
coupled by relatively weak capacitances has been considered 
as a possible physical implementation of such
objects. Numerical experiments have shown that a 
time-monochromatic external force applied to several
oscillators leads to the formation of long-lived and 
nontrivially interacting vortices in the system against 
a quasistationary background in a wide range of parameters.
The dynamics of vortices depends on the method
of ``coupling'' of the opposite sides of a rectangular array
by links, which determines the topology of the
resulting manifold (torus, Klein bottle, projective plane,
M\"obius strip, ring, or disk).
\end{abstract}

\maketitle

\section {Introduction}

As known, nonlinear complex wavefields can form
quantum vortices in two- and three-dimensional
spaces [1-7]. A typical example is vortices in trapped
Bose condensates of cold atoms (which are described
by the condensate wavefunction $\Psi({\bf r},t)$ within the
Gross-Pitaevskii equation). Nontrivial dynamic
properties of these objects attract attention of
researchers (see, e.g., [8-19]). Vortex structures exist
not only in continuous media but also in discrete systems 
(vortices and vortex solitons on lattices; see [20-29] 
and references therein). Dynamic systems on lattices
allow different physical and mathematical realizations. 
In particular, vortices in arrays of coupled
nonlinear oscillators can exist because each site
involves the canonical complex order parameter
$a_n=\sqrt{S_n}\exp(i\Theta_n)=A_n(t)\exp(-i\omega_0 t)$ , where $S_n$ and
$\Theta_n$ are the action and angle variables for a single oscillator, 
respectively, and $\omega_0$ is the frequency of oscillations 
in the limit of small amplitudes. The phase $\Theta$
changing slowly upon the bypass along a closed contour
along the links can acquire an increment multiple
of $2\pi$, thus forming a discrete vortex. However, for
such an object to have a pronounced localized core
and to be long-lived against a modulationally stable
nonzero ``density'' background $S$, the effect of nonlinearity 
should be defocusing. The core of the vortex
(dip of the density) can be in this case as narrow as
about one step of the lattice and even smaller in a certain 
sense, but the effect of its phase covers the entire
system. This property distinguishes ordinary vortices
from localized vortex solitons occurring upon focused
nonlinearity. The interaction of discrete vortices with
each other and with sites of the lattice is responsible 
for the complex dynamics, which is the subject of this work.

One of the relatively simple and universal mathematical 
models allowing vortex solutions is the weakly dissipative 
discrete nonlinear Schr\"odinger equation with pump
\begin{eqnarray}
&&i(\dot A_n+\gamma\omega_0 A_n) = g|A_n|^2A_n \nonumber\\
&&\qquad\qquad+\frac{1}{2}\sum_{n'}c_{n,n'}(A_n-A_{n'}) +f_n(t),
\label{A_n_eq}
\end{eqnarray}
where $\gamma$  is a small linear damping coefficient, $g$ is a
nonlinear coefficient, $c_{n,n'}$ is a (real) coupling
matrix, and $f_n(t)$ is the complex envelope of an external 
quasimonochromatic force (near the resonance
frequency). In particular, various metamaterials are
described by an equation of this type (see, e.g., [30]
and references). It summarizes effects of nonlinearity,
dispersion, dissipation, and resonance pumping. Real
possible oscillators satisfy such a universal equation
only approximately and only in a weakly nonlinear
regime. Consequently, Eq.(1) is insufficient for studying 
the possibility of fabrication of artificial materials
that can demonstrate discrete vortices, and it is of interest 
to study strongly nonlinear physically implementable systems.

It is noteworthy that electric circuits with reverse-biased 
varactor diodes (variable capacitances depending on
the applied voltage) are practically convenient implementations 
of coupled nonlinear oscillators [31-45].
In particular, they were used to perform experiments
simulating the dynamics of solitons on an integrable
Toda chain [31-34]. In addition to diodes, nonlinear
capacitors based on special dielectric films have also
been developed [46,47].

Depending on the design of a circuit, nonlinearity
can be both focusing at large scales and defocusing.
Until now, focusing variants characterized by the
modulation instability of long waves have primarily
been studied for discrete solitons, breathers, and vortex 
solitons [20,22,42,43]. These strongly localized
structures are formed on modulationally unstable networks 
even with small sizes (smaller than $10\times 10$) and
occupy only a few cells [43]. Unlike them, the reliable
observation of vortices on modulationally stable arrays
of coupled oscillators requires a large number of elements,
at least $30\times 30$, because the spatial distribution
of the phases of vortices is usually delocalized. The
fabrication of such large arrays requires high costs.
This is likely why vortices on networks have not yet
been observed experimentally. An interesting fact that
systems of electric oscillators allow such design of couplings 
that provides topologically nontrivial discrete
manifolds, e.g., M\"obius strip, Klein bottle, and projective 
plane, still does not attract attention. Since vortices 
are long-range objects, the topology of manifolds
should strongly affect them. As far as I know, the
dynamics of discrete vortices on complex manifolds
has not yet been studied (in contrast to discrete solitons 
on a quasi-one-dimensional M\"obius strip [48]).
The aim of this work is to partially fill this gap by only
the numerical simulation of a fairly realistic electric
model. As will be seen below, already the first results
appear to be nontrivial and interesting. In particular,
long-lived vortices on a discrete weakly dissipative
electric circuit under a periodic external action in the
nonlinear resonance regime are observed in a numerical 
experiment for the first time.

\begin{figure}
\begin{center}
\epsfig{file=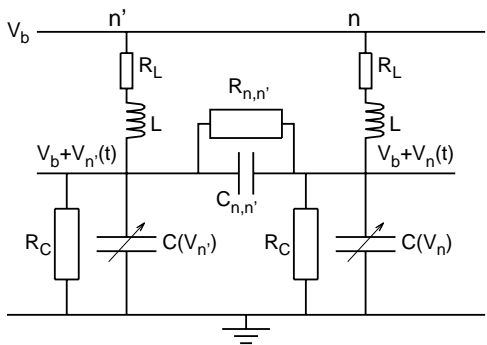, width=70mm}
\end{center}
\caption{Circuit of coupled nonlinear electric oscillators.
Only a fragment of the entire circuit (two cells and a link
between them) is shown.
}
\label{scheme} 
\end{figure}

\section{Description of the model}

Figure 1 shows the considered circuit consisting of
nonlinear electric oscillators coupled by relatively
weak capacitive couplings between them. The relation
between this circuit and Eq.(1) was recently discussed
in [45]. A state of the system is described by voltages
$V_n(t)$ and by currents $I_n(t)$ through inductors $L$ toward
the corresponding capacitor. The use of toroidal coils
is suggested in order to exclude mutual induction.
Here, capacitors $C(V_n)$ are nonlinear elements. For
simplicity, the voltage dependence of the capacitance
is used in the form
\begin{equation}
C(V_n)=C_0(1+V_n^2/V_*^2),
\label{CV_symm}
\end{equation}
where $V_*$ is a parameter of about several volts. This
symmetric dependence is characteristic of capacitors
with dielectric films [46, 47]. The application of the
bias voltage $V_b$ makes it possible to use also varicaps,
which (in parallel connection to a normal capacitor)
are approximately described by the formula (see, e.g.,[49])
\begin{equation}
C(V_n)=C_0\Big[\mu+(1-\mu)/(1+V_n/V_*)^\nu\Big],
\label{CV_diode}
\end{equation}
where $0<\mu<1$ is a parameter taking into account
the normal capacitor connected in parallel and $\nu$ is a
fitting parameter of the diode, which depends on the
technology of fabrication and is usually in the range of
$0.3\lesssim \nu\lesssim 6.0$ ($\nu=1$ should be taken to obtain a Toda
chain). In any case, the stored electrostatic energy on
the capacitor (additional to that in the state $V_n=0$ ) is
determined by the formula
\begin{equation}
W(V_n)={\textstyle\int}_{0}^{V_n}C(u)u du,
\end{equation}
and the additional electric charge is
\begin{equation}
q_n=q(V_n)={\textstyle\int}_{0}^{V_n}C(u) du.
\end{equation}

All capacitances $C_{n,n'}$ on couplings are assumed to
be linear and small compared to $C_0$. Generally speaking, 
couplings are not necessarily identical, which
allows an additional degree of freedom in the design of
networks with spatially inhomogeneous properties. It
is possible to fabricate locally periodic, quasicrystalline, 
and (pseudo)random lattices.

Under the assumption that connecting wires have
negligibly small resistances, inductances, and capacitances,
the real spatial arrangement of the elements of
the circuit is insignificant. Circuits with almost any
topology are obviously implementable because of the
flexibility and small cross section of wires.

For the electric model to be realistic, it should
include dissipative elements such as a low active resistance 
of the coil $R_L\ll\sqrt{L/C_0}$ and high leakage resistances 
of capacitors $R_C\gg\sqrt{L/C_0}$ and $R_{n,n'}\gg\sqrt{L/C_0}$.
For simplicity, a possible nonlinear dependence of
$R_C$ on the voltage applied to the capacitor is ignored.
Furthermore, to compensate energy losses in the system, 
an alternating voltage ${\cal E}_n(t)$ (generator connected
in series to the coil is not shown in Fig.1) is applied to
a few oscillators.

In the general case, the steady-state voltage on the
capacitor is slightly different from $V_b$ because of the
finiteness of $R_C$, but this difference can be neglected in
the measure of a very small parameter  $R_L/R_C$. Then,
the system of equations of motion has the form
\begin{eqnarray}
&&C(V_n)\dot V_n+\sum_{n'} \big[C_{n,n'}(\dot V_n -\dot V_{n'})\nonumber\\
&&\qquad\qquad+(V_n -V_{n'})/R_{n,n'}\big]+V_n/R_C= I_n,
\label{current}\\
&&L \dot I_n + V_n + R_L I_n = {\cal E}_n(t).
\label{voltage}
\end{eqnarray}
To solve this system with respect to the time derivatives
$\dot V_n$ in the numerical algorithm, it is possible to use 
iterations corresponding to the Euler scheme for relaxation to
the equilibrium position. In this case, the convergence of
iterations is ensured by the symmetry and positive definiteness 
of the matrix $\{[C(V_n)+\sum_m C_{n,m}]\delta_{n,n'}-C_{n,n'}\}$
appearing in Eq.(6). The result of iterations is then
substituted into a 4th-order Runge-Kutta algorithm 
for time evolution. The calculations are performed 
in dimensionless variables $L=1$, $C_0=1$, and $V_*=1$. 
Correspondingly, the frequency and period of
small oscillations are $\omega_0=1$ and $T_0=2\pi$ , respectively.
It is noteworthy that the energy $\varepsilon_n=LI_n^2/2+W(V_n)$
of each oscillator in the absence of couplings and dissipation 
would be conserved.

The requirement that the nonlinear frequency shift 
$\delta\omega_{\rm nl}=g|A|^2$
of a separate oscillator be negative is fundamentally 
important because, as shown in [45], the
dominant elements of the matrix $c_{n,n'}$ of the corresponding 
equation (1) are negative, although the
energy of couplings in our model is positive definite. It
is particularly easy to test this by the example of an
infinite square lattice by calculating the dispersion law
for the linearized system and verifying that the quadratic 
correction to the frequency at low wavenumbers
is negative. When the signs of the nonlinear coefficient
and dispersion correction are the same, the action of
nonlinearity will be defocusing in the quasi-continuum 
limit, which is necessary for the formation of a
stable background on which vortices exist. This condition 
with the function (2) is satisfied automatically
(because $g=-1/4$ in this case) and with the function 
(3) is satisfied in the region of the parameters
\begin{equation}
g=\nu(1-\mu)[-3+\nu(1-4\mu)]/24<0. 
\end{equation}

It is assumed that dissipation in the coil prevails over leakage 
dissipation, so that the Q-factor of a separate oscillator
\begin{equation}
Q^{-1}=\gamma=\left(R_L\sqrt{C_0/L}+R_C^{-1}\sqrt{L/C_0}\right)/2
\end{equation}
is determined primarily by the first term. The numerical
experiments show that a high Q-factor $Q\gtrsim 10^{3}$ is
required for the observation of vortices. This requirement 
seems quite realistic. In particular, for the coil
with the inductance $L=1.0\times 10^{-4}$ H and resistance $R_L=1 \,\Omega$
(such a coil made of a wire with a length of 10 m
and a cross section of 0.2 mm$^2$ has dimensions of several 
centimeters) at the values $C_0=1.0\times 10^{-10}$ F and 
$R _C>10^7\,\Omega$, we obtain $\omega_0=1.0\times 10^7$ rad/s 
(corresponding to a frequency of about 1.6 MHz) and a high Q-factor
$Q>10^3$. The resistance of copper decreases at low
temperatures and the Q-factor increases.

\begin{figure}
\begin{center}
\epsfig{file=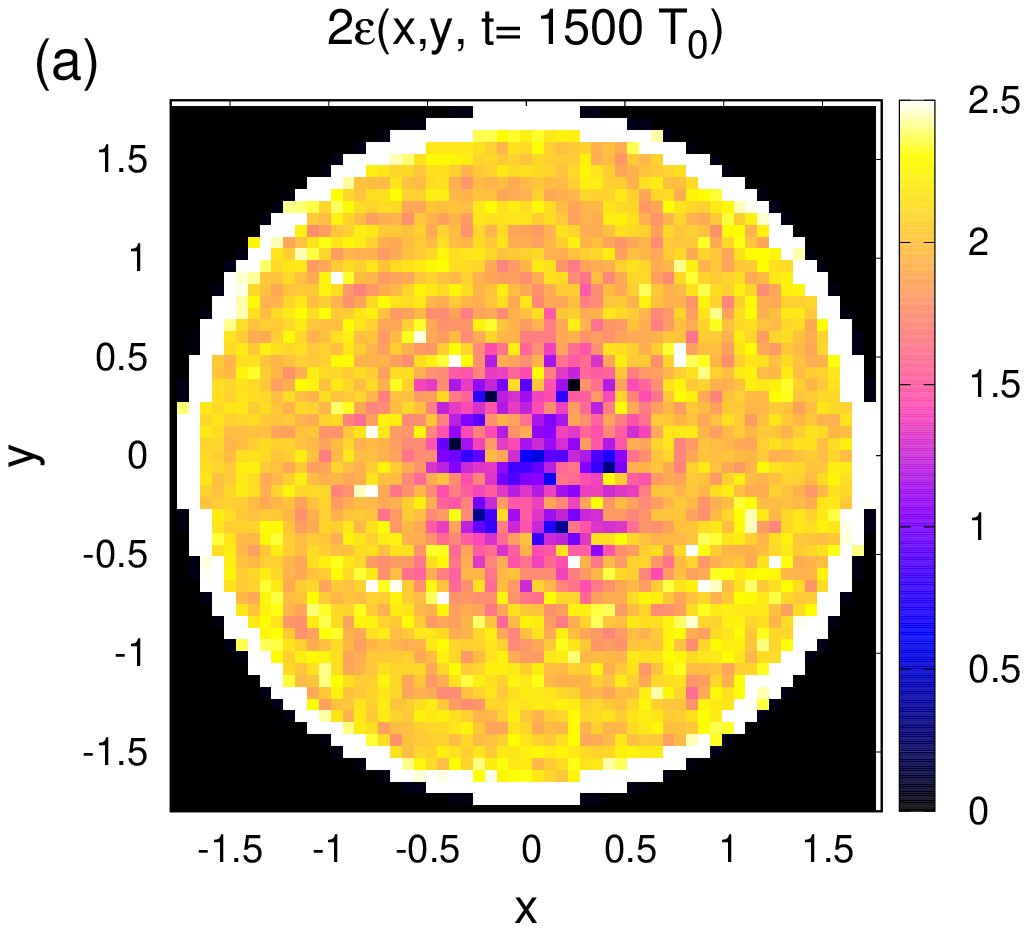, width=70mm}
\epsfig{file=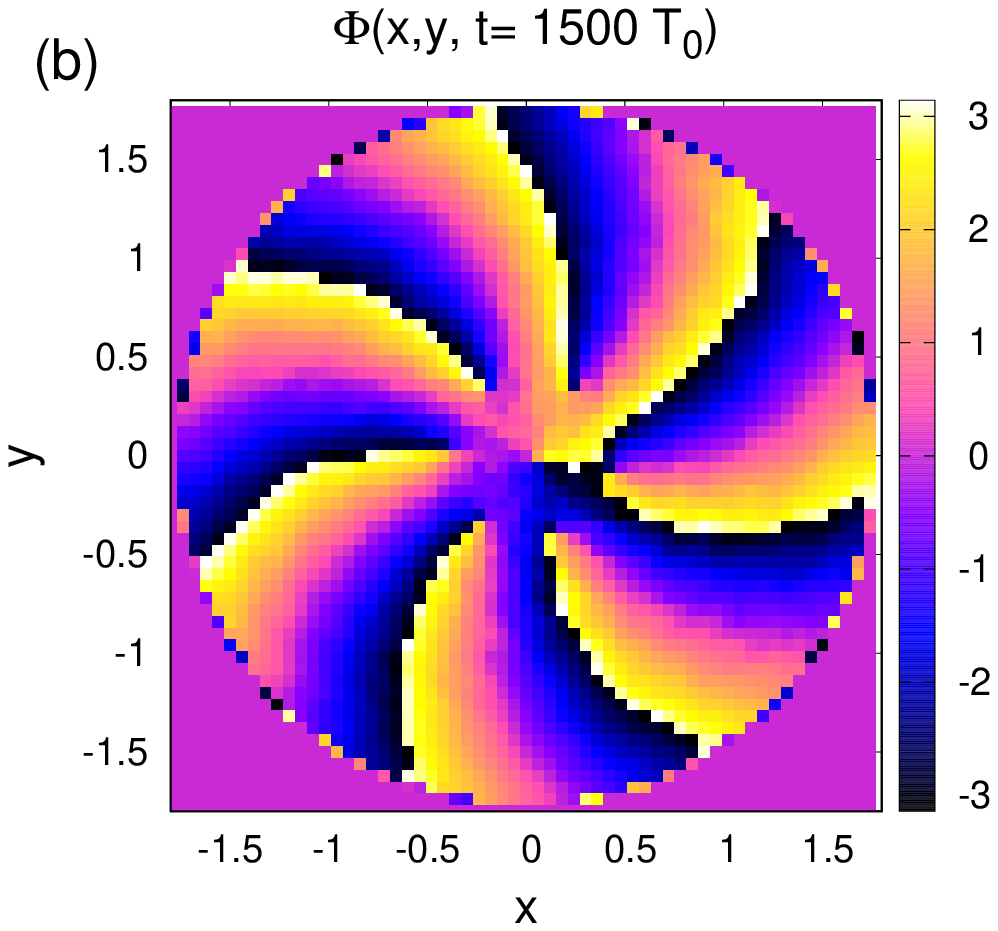, width=70mm}
\end{center}
\caption{Maps of (a) the doubled energy of
oscillators on the lattice and (b) the phase for the rotating
cluster of seven vortices on the disk.
}
\label{disc} 
\end{figure}
\begin{figure}
\begin{center}
\epsfig{file=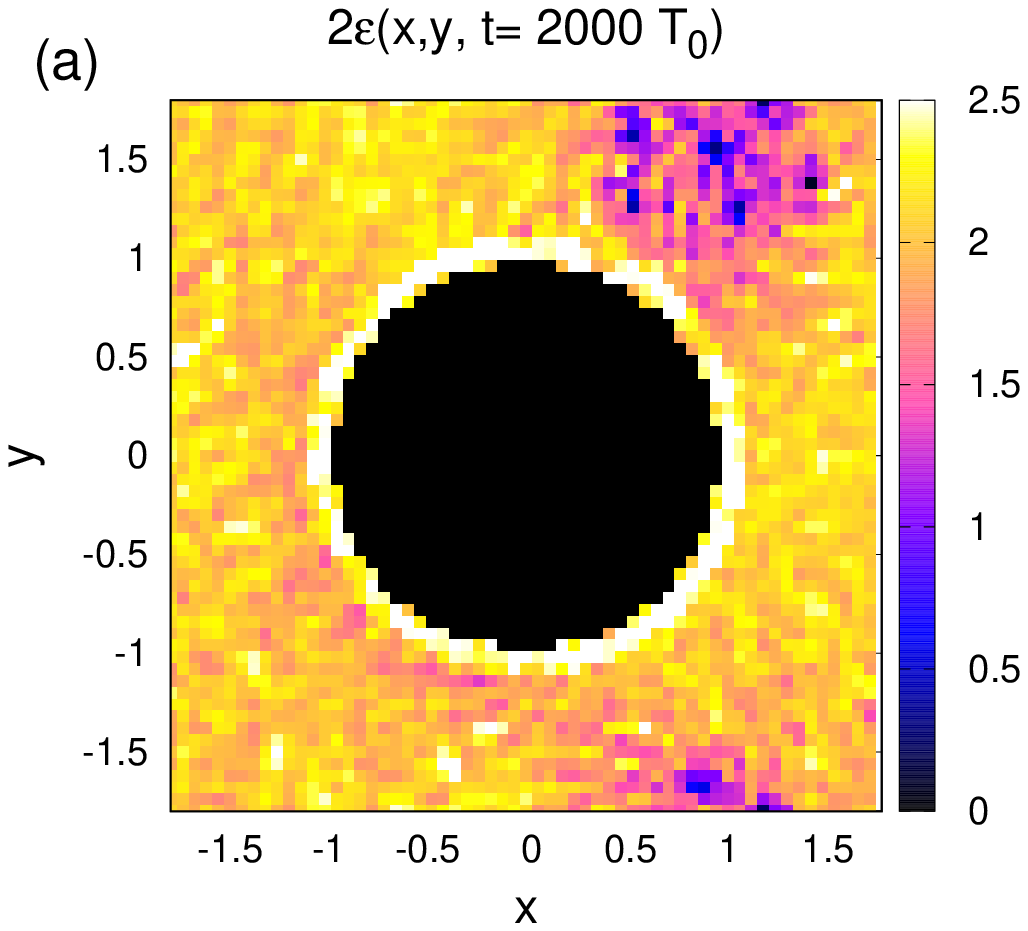, width=70mm}
\epsfig{file=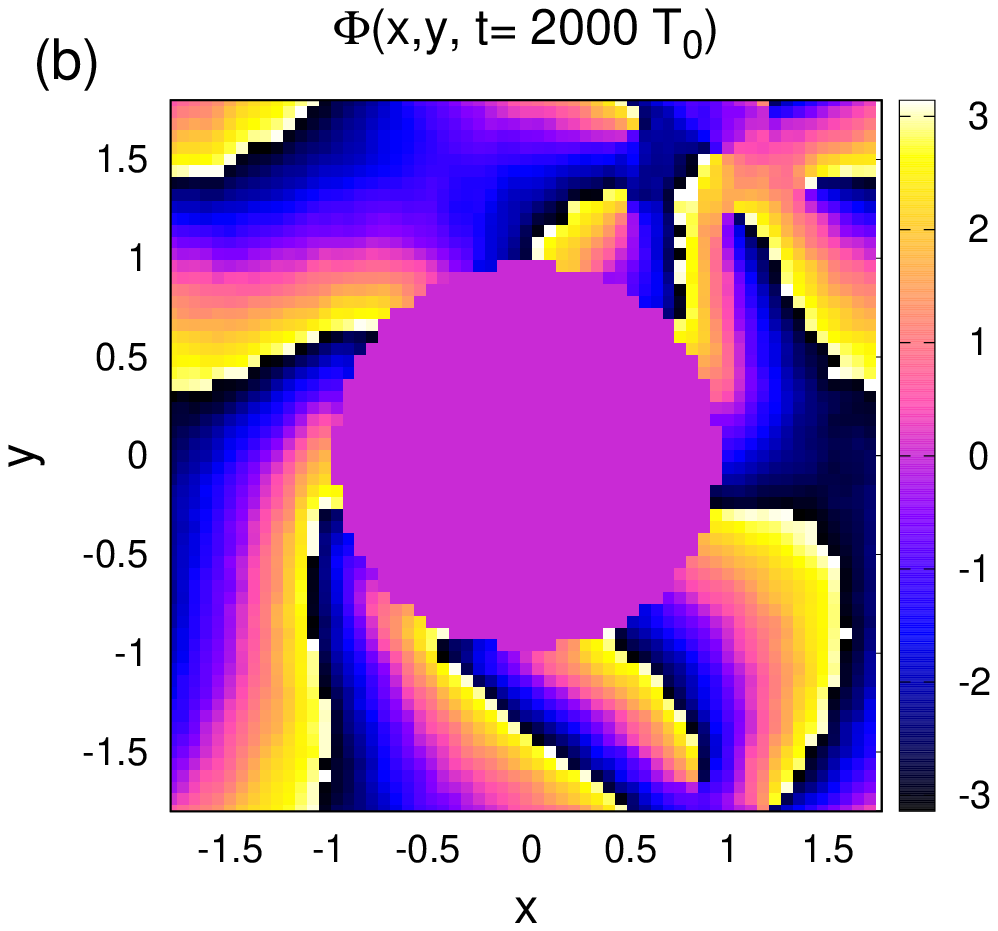, width=70mm}
\end{center}
\caption{Maps of the (a) doubled energy and
(b) phase for the rotating cluster of seven vortices on the
torus with the hole.
}
\label{torus} 
\end{figure}
\begin{figure}
\begin{center}
\epsfig{file=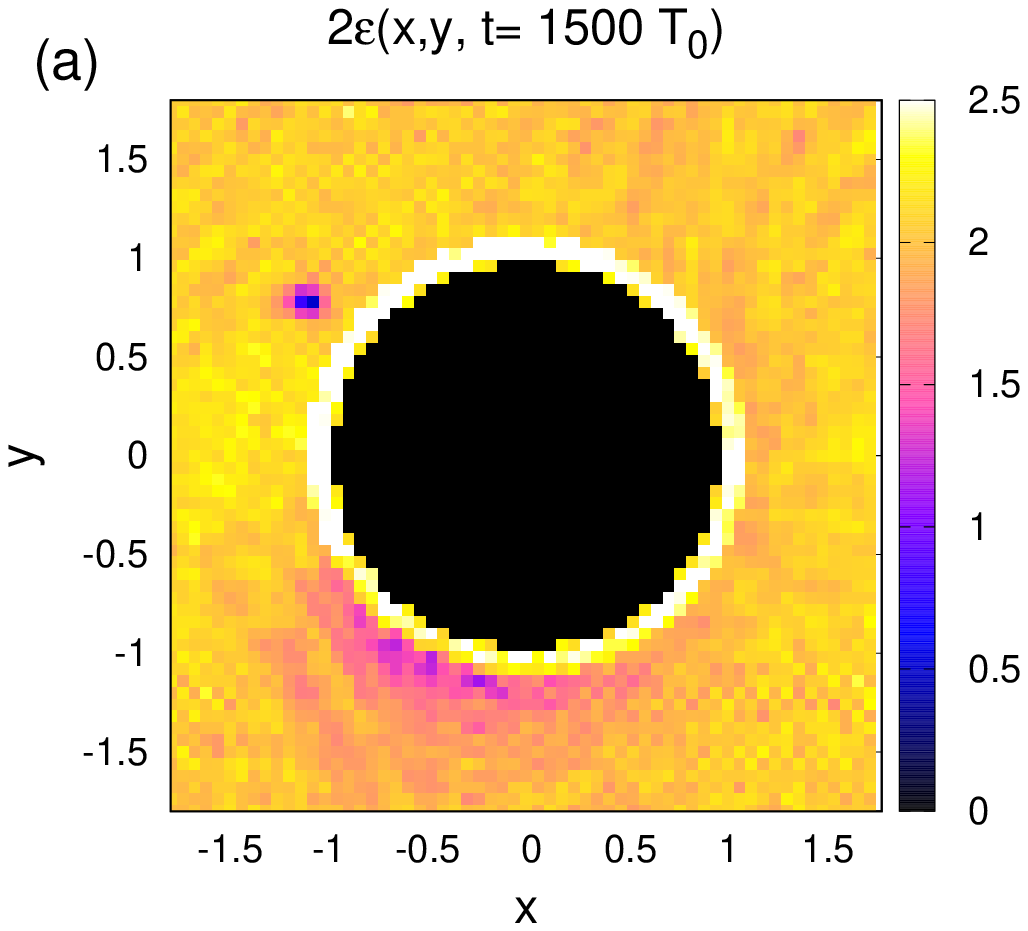, width=70mm}
\epsfig{file=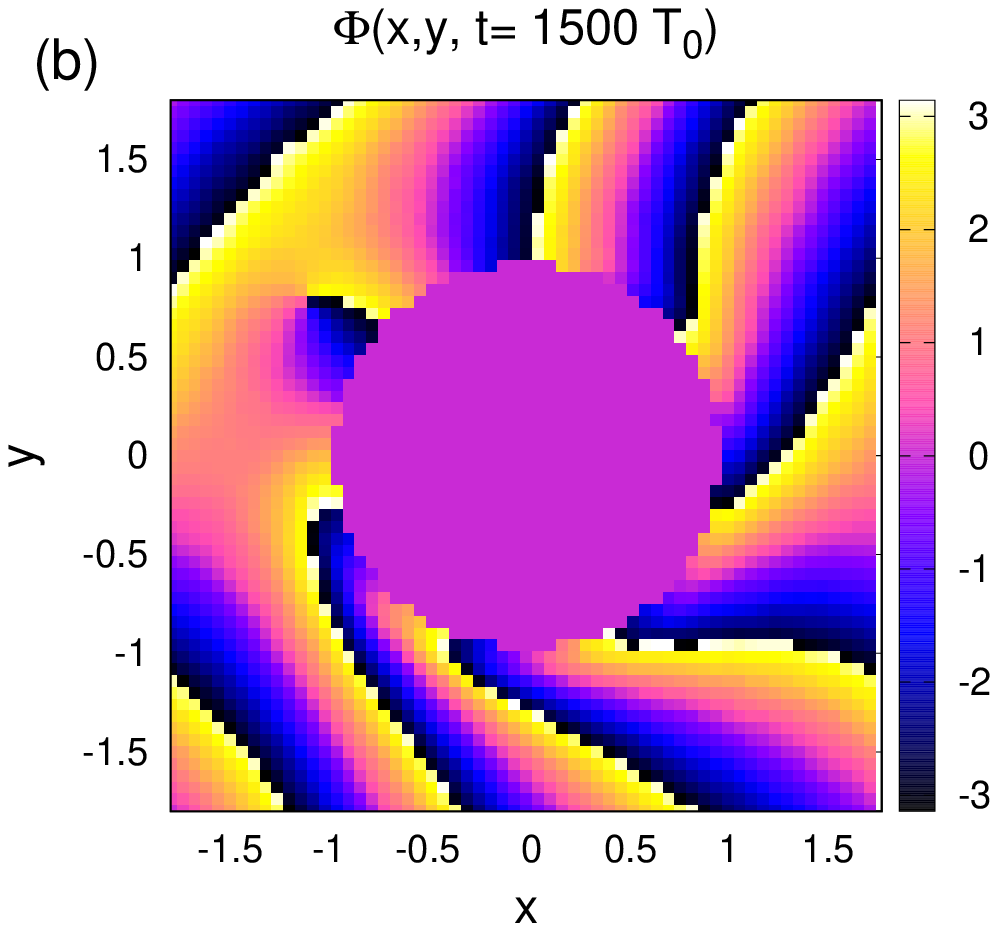, width=70mm}
\end{center}
\caption{Maps of the (a) doubled energy and
(b) phase for the last remaining vortex on the Klein bottle
with the hole.
}
\label{Klein} 
\end{figure}
\begin{figure}
\begin{center}
\epsfig{file=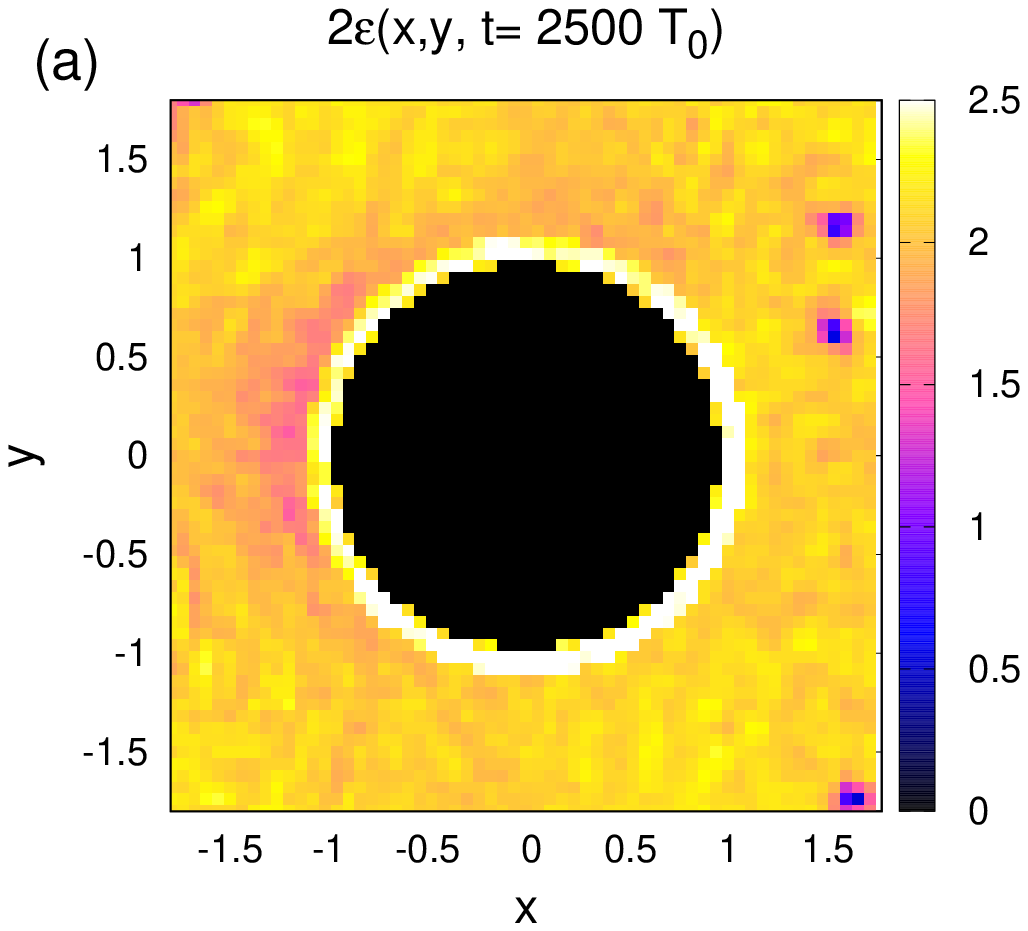, width=70mm}
\epsfig{file=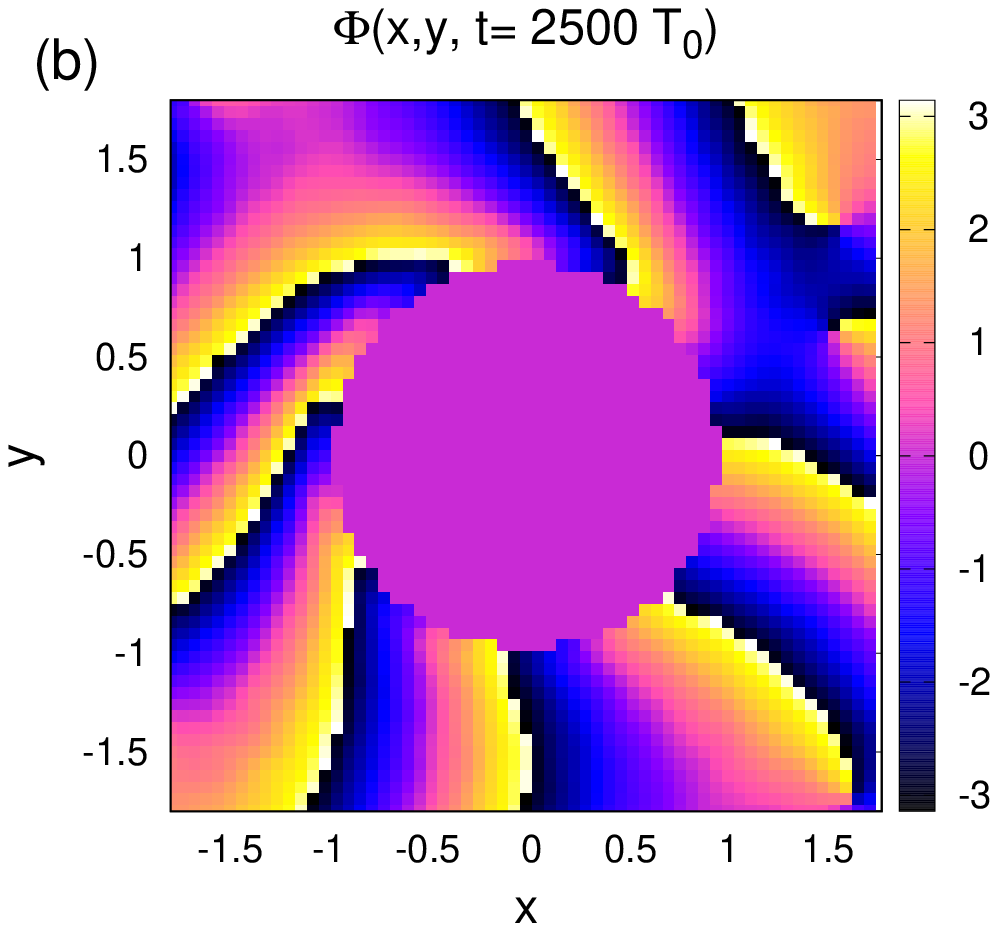, width=70mm}
\end{center}
\caption{Maps of the (a) doubled energy and
(b) phase for three vortices on the projective plane with the
hole that move along the perimeter of the square. Finally,
only one vortex remains.
}
\label{projp} 
\end{figure}
\begin{figure}
\begin{center}
\epsfig{file=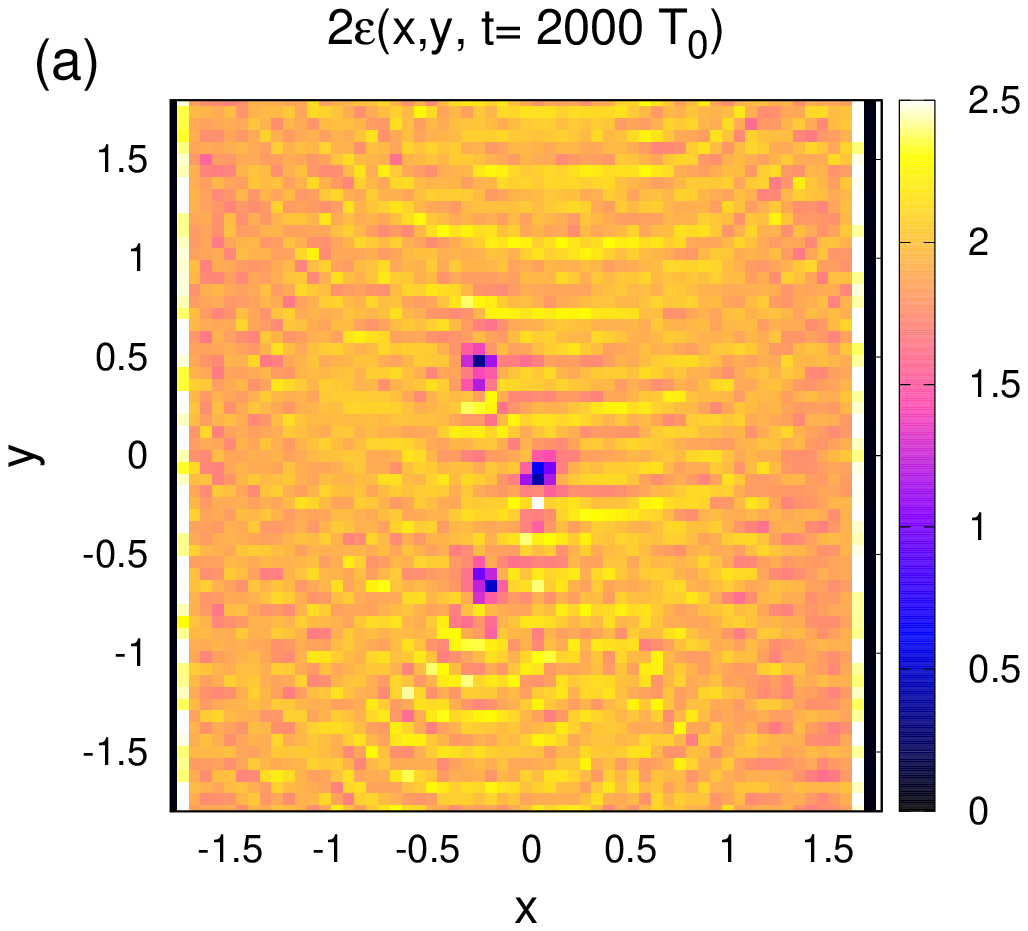, width=70mm}
\epsfig{file=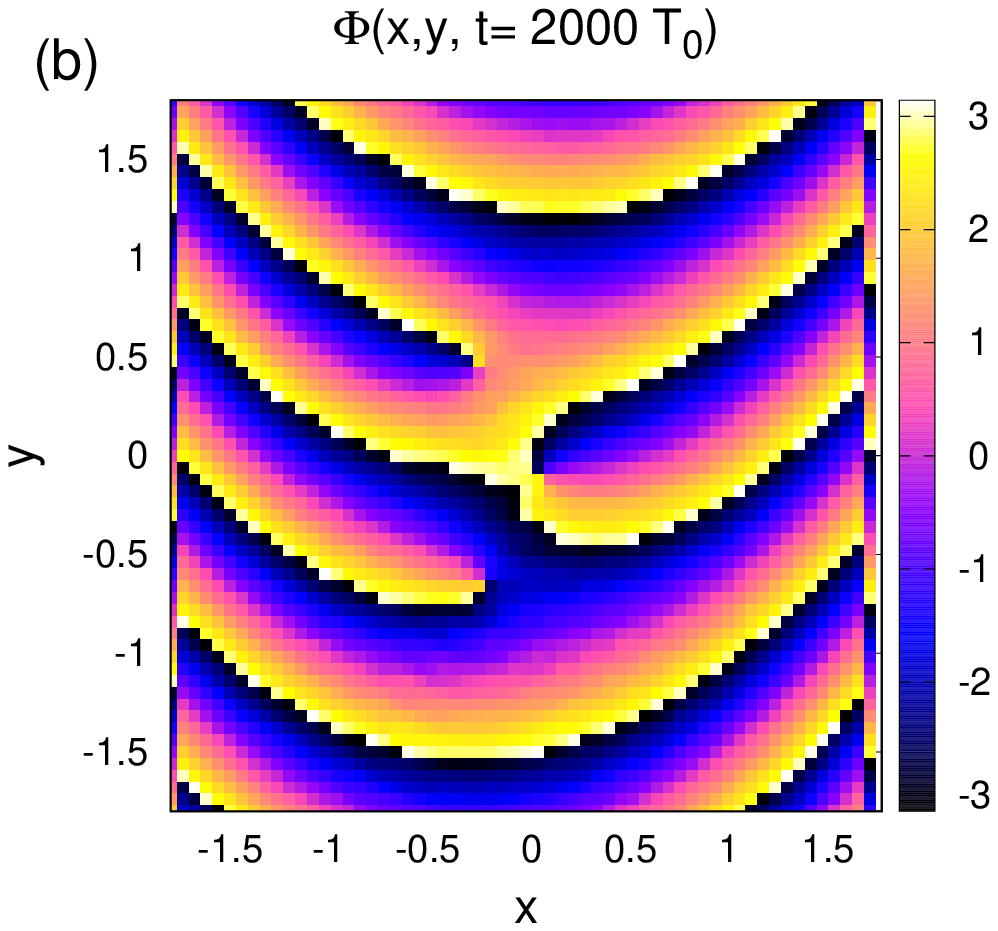, width=70mm}
\end{center}
\caption{Maps of the (a) doubled energy and
(b) phase for vortices on the M\"obius strip that move in parallel 
to the boundary.
}
\label{Mobius} 
\end{figure}

\section{Numerical  experiments}

In the computer calculations, an element of a two-dimensional 
array $N_1\times N_2$ is specified by the multiindex 
$n=n_2+N_2n_1$, where $0\leq n_1\leq N_1-1$ and
$0\leq n_2\leq N_2-1$. Only couplings between nearest
neighbors or between elements on the opposite sides
of the rectangle can be nonzero. If both pairs of the
opposite sides are joined by couplings in the direct order 
[$(0,n_2)\leftrightarrow (N_1-1,n_2)$, $(n_1,0)\leftrightarrow (n_1,N_2-1)$],
the resulting manifold is a torus. If one pair of the
opposite sides are joined in the direct order and the
second pair are joined in the reverse order
[$(n_1,0)\leftrightarrow (N_1-1-n_1,N_2-1)$], the resulting manifold
is a Klein bottle (on this non-orientable manifold, the
transition of a vortex through the edge along the
``reversed'' direction results in its appearance with the
opposite sign near the opposite side of the rectangle at
the point symmetric with respect to the center). If
both pairs of opposite sides are joined in the reverse
order, the resulting manifold is a projective plane.
Furthermore, if the first pair of opposite sides are not
joined altogether, whereas the second pair of opposite
sides are joined in the reverse order, the resulting manifold 
is a M\"obius strip. Vortices on the disk were also
simulated.

In the numerical experiments reported below,
$N_1=N_2=60$, and, for convenience, sites were
located on the square lattice with an arbitrarily chosen
step of $h=0.06$ for the total size of the system to be
about unity. The external signal ${\cal E}_n(t)$ was taken to be
monochromatic with the frequency $(1-\Delta)$ for the system 
to be in the state of nonlinear resonance, where
the amplitude of oscillations is determined primarily
by the frequency of pumping according to the estimate
$gS\approx-\Delta$ and, to a smaller extent, by the amplitude of
pumping. For the disk and M\"obius strip, the pumping was
applied to sites located along the natural boundary. A
disk was cut from the torus, Klein bottle, and projective 
plane, and the signal was applied on sites near the
formed boundary. For the formation of vortices to be
energetically favorable, the initial phases of the signal
depended on $n$, uniformly increasing at bypass along
the boundary to the value $2\pi M$, where $M$ is an integer
about 10. In the case of the disk and torus, the number
$M$ determined the number of vortices in the formed
cluster (at the sufficient amplitude of force). In all
other cases, no simple correspondence existed
between $M$ and the number of vortices: they appeared
and disappeared both in the form of vortex-antivortex
pairs and singly at collisions with the circular boundary.

Because of a large number of parameters involved
in numerical experiments, the results obtained are still
insufficient for the complete description of all possible
dynamic regimes. For this reason, only some examples
are given. However, even these examples clearly
demonstrate nontrivial properties of vortices on electric 
circuits. The parameters were as follows: 
$R_L=0.001$, $1/R_C=0.0001$, $C_{n,n'}=0.1$, $1/(R_{n,n'}C_{n,n'})=0.0001$, 
the amplitude of the external signal was $0.06$,
detuning of its frequency $\Delta=0.14$, and $M=7$. All
oscillators in the initial state had $I_n(0)=0$ and the
same values $V_n(0)\sim 1$. After an initial transient period
continuing about $1000T_0$, a quite uniform background
density was established on which nucleated vortices
moved. Their further evolution significantly depended
on the topology of the manifold, as seen in Figs.2-6,
which show maps of the energies of oscillators $\varepsilon_n$ on
the lattice, as well as the quantities $\Phi_n=\arctan(I_n/V_n)$,
which are qualitatively similar to
canonical phases $\Theta_n$. In particular, a vortex cluster was
formed on the disk, as seen in Fig.2. Vortices in this
cluster rotated slowly and changed their mutual
arrangement. A similar quiescent regime of rotation of
the cluster also occurred in the case of the torus shown
in Fig.3. The dynamics on the Klein bottle and projective 
plane occurred differently, where compact
clusters were not formed and the number of vortices
decreased gradually with the time (see Figs. 4 and 5,
where the difference between the boundary conditions
is due to the direct or reverse order of joining of opposite 
sides and becomes obvious from the comparison
of maps of the ``phase'' $\Phi$). The last remaining vortex
on the Klein bottle ``was trapped'' near the edge of the
hole. The last vortex on the projective plane moved
along the perimeter of the square clockwise. Vortices
on the M\"obius strip moved in parallel to the boundary,
rarely annihilating pairwise at collisions. In this case,
three vortices shown in Fig.6 still remained at times to
$5000T_0$.

Calculations were also performed with other sets of
parameters, e.g., with increased or decreased active
resistances, with different frequencies and amplitudes
of pumping, as well as with nonuniform profiles of
couplings (different $C_{n,n'}$ values in different parts of a
manifold). Generally speaking, the dynamics of vortices 
was sometimes strongly different from that
described above. In some cases, dark solitons were
observed in addition to vortices. All details cannot be
discussed here. Important is the fact that long-lived
vortices are not exclusive objects requiring fine adjustment
of the system but occur in a wide region of parameters.

\section{Conclusions}

To summarize, a new example of a dynamic system
describing modulationally stable nonlinear waves in a
weakly dissipative two-dimensional electric circuit has
been presented. The numerical simulation has confirmed 
a scenario according to which resonance
pumping concentrated at the edge of such systems
results in the formation of an almost uniform background 
density against which discrete vortices appear
and exist for a long time. Such vortices on topologically 
complex manifolds have been numerically
demonstrated for the first time. The derivation (in
spatially continuous limit) of equations of motion
directly for coordinates of ``point'' vortices in the presence 
of dissipation and pump is a problem for the
future. Such equations will apparently be more complex 
than those for the case of free relaxation considered 
in [29,45].

This work was motivated in particular by interest in
the development of artificial materials (including
three-dimensional) that can support macroscopic
quantum vortices without the Planck constant at room
temperature. It is clear that such a material consisting
of many millions of units cannot be constructed from
radio engineering elements. Its elements should be
simpler and cheaper. However, I hope that the general
idea of this work will be practically useful for further
studies in this field.

\end{document}